\documentclass[twocolumn]{aastex63}

\usepackage{amsmath,amssymb,amsfonts,bm}
\usepackage{epstopdf}
\usepackage{epsfig}
\usepackage{natbib,caption2}
\usepackage{graphicx}   
\usepackage{float}
\usepackage{longtable}
\usepackage{graphics}
\usepackage{hyperref}
\usepackage{color}
\usepackage{calc}
\usepackage{threeparttable}

\newcommand \beq{\begin{equation}}
\newcommand \eeq{\end{equation}}
\newcommand \bey{\begin{eqnarray}}
\newcommand \eey{\end{eqnarray}}

\newcommand \kpc{\, {\rm kpc} }

\newcommand \SA {\rm{\tau_{\rm L}}} 
\newcommand \SAM {\rm{\tau_{\rm M}}} 
\newcommand \SARE {\rm{\tau_{\rm L, R_{e}}}}
\newcommand \SAC {\rm{\tau_{\rm L, cen}}}
\newcommand \SZ {[Z/{\rm H}]_{\rm L}}
\newcommand \SZM {[Z/{\rm H}]_{\rm M}}
\newcommand \SZRE {[Z/{\rm H}]_{\rm L, R_{e}}}
\newcommand \SZC {[Z/{\rm H}]_{\rm L, cen}}
\newcommand \SDRE {\rm{D4000_{\rm R_{e}}}}
\newcommand \SDC {\rm{D4000_{\rm cen}}}
\newcommand \grad {\nabla}
\newcommand \GD {\grad{\rm{D4000}}}
\newcommand \GA {\grad{\tau_{\rm L}}}
\newcommand \GZ {\grad{[Z/{\rm H}]_{\rm L}}}
\newcommand \GAM {\grad{\tau_{\rm M}}}
\newcommand \GZM {\grad{[Z/{\rm H}]_{\rm M}}}
\newcommand \Msun{M_\odot} 
\newcommand \Msunyr {M_\odot~{\rm yr{^{-1}}}}

\newcommand \Ms{M_{\star}}

\newcommand{\Reff}{R_{\rm{e}}}

\newcommand{\Sc}{\Sigma_{1\rm{kpc}}}
\newcommand{\DSc}{\Delta \log\Sigma_{1\rm{kpc}}}

\newcommand{\gsim}{\lower.5ex\hbox{$\; \buildrel > \over \sim \;$}}
\newcommand{\lsim}{\lower.5ex\hbox{$\; \buildrel < \over \sim \;$}}
\newcommand{\ha}{\hbox{H$\alpha$}}

\def\correspondingauthors#1{{
\renewcommand\thefootnote{\hskip-12pt}
\footnote{Corresponding authors: #1\ifmodern\vrule depth 5pt
width 0pt\relax\fi}}}

\submitjournal{ApJ}

\shorttitle{\space}
\shortauthors{Chen et al.}

\begin{document}

\title{The Most Predictive Physical Properties for the Stellar Population Radial Profiles of Nearby Galaxies}

\correspondingauthors{Hong-Xin Zhang, Xu Kong}

\author{Guangwen Chen}
\affiliation{CAS Key Laboratory for Research in Galaxies and Cosmology, Department of Astronomy, University of Science and Technology of China, Hefei 230026, China\\}
\affiliation{School of Astronomy and Space Science, University of Science and Technology of China, Hefei 230026, China}
\email{guangwen@mail.ustc.edu.cn \\ hzhang18@ustc.edu.cn \\ xkong@ustc.edu.cn}

\author{Hong-Xin Zhang}
\affiliation{CAS Key Laboratory for Research in Galaxies and Cosmology, Department of Astronomy, University of Science and Technology of China, Hefei 230026, China\\}
\affiliation{School of Astronomy and Space Science, University of Science and Technology of China, Hefei 230026, China}

\author{Xu Kong}
\affiliation{CAS Key Laboratory for Research in Galaxies and Cosmology, Department of Astronomy, University of Science and Technology of China, Hefei 230026, China\\}
\affiliation{School of Astronomy and Space Science, University of Science and Technology of China, Hefei 230026, China}

\author{Zesen Lin}
\affiliation{CAS Key Laboratory for Research in Galaxies and Cosmology, Department of Astronomy, University of Science and Technology of China, Hefei 230026, China\\}
\affiliation{School of Astronomy and Space Science, University of Science and Technology of China, Hefei 230026, China}

\author{Zhixiong Liang}
\affiliation{CAS Key Laboratory for Research in Galaxies and Cosmology, Department of Astronomy, University of Science and Technology of China, Hefei 230026, China\\}
\affiliation{School of Astronomy and Space Science, University of Science and Technology of China, Hefei 230026, China}

\author{Xinkai Chen}
\affiliation{CAS Key Laboratory for Research in Galaxies and Cosmology, Department of Astronomy, University of Science and Technology of China, Hefei 230026, China\\}
\affiliation{School of Astronomy and Space Science, University of Science and Technology of China, Hefei 230026, China}

\author{Zuyi Chen}
\affiliation{CAS Key Laboratory for Research in Galaxies and Cosmology, Department of Astronomy, University of Science and Technology of China, Hefei 230026, China\\}
\affiliation{School of Astronomy and Space Science, University of Science and Technology of China, Hefei 230026, China}

\author{Zhiyuan Song}
\affiliation{CAS Key Laboratory for Research in Galaxies and Cosmology, Department of Astronomy, University of Science and Technology of China, Hefei 230026, China\\}
\affiliation{School of Astronomy and Space Science, University of Science and Technology of China, Hefei 230026, China}




\begin{abstract}

We present a study on the radial profiles of the D4000, luminosity-weighted stellar ages $\SA$, and luminosity-weighted stellar metallicities $\SZ$ of 3654 nearby galaxies ($0.01 < z < 0.15$) 
using the IFU spectroscopic data from the MaNGA survey available in the SDSS DR15, in an effort to explore the connection between median stellar population 
radial gradients (i.e.,\ $\GD$, $\GA$, $\GZ$) out to $\sim$ 1.5 $R_{e}$ and various galaxy properties, including stellar mass ($\Ms$), specific star formation rate 
(sSFR), morphologies, and local environment.\ We find that $\Ms$ is the single most predictive physical property for $\GD$ and $\GZ$.\ The most predictive 
properties for $\GA$ are sSFR and, to a lesser degree, $\Ms$.\ The environmental parameters, including local galaxy overdensities and central--satellite division, 
have virtually no correlation with stellar population radial profiles for the whole sample, but the $\GD$ of star-forming satellite galaxies with $\Ms$ $\lesssim$ 10$^{10}~\Msun$ 
exhibit a significant positive correlation with galaxy overdensities.\
Galaxies with lower sSFR have on average steeper negative stellar population gradients, and this sSFR dependence is stronger for more massive star-forming galaxies.\
The negative correlation between the median stellar population gradients and $\Ms$ are best described largely as segmented relationships, whereby median  
gradients of galaxies with $\log\Ms \lesssim 10.0$ (with the exact value depending on sSFR) have much weaker mass dependence than galaxies with higher 
$\Ms$.\ While the dependence of the radial gradients of ages and metallicities on T-Types and central stellar mass surface densities are generally not significant, 
galaxies with later T-Types or lower central mass densities tend to have significantly lower D4000, younger $\SA$, and lower $\SZ$ across the radial ranges 
probed in this study.\

\end{abstract}



\section{Introduction}
\label{sec:intro}

The stellar population distribution within galaxies holds important clues to the assembly history of galaxies.\
In the classical dissipative collapse models of galaxy formation (e.g., \citealt{Eggen_Lynden_Sandage_1962,Larson1974,Carlberg1984,Pipino2010}), 
the outer parts of galaxies have less efficient star formation and self-enrichment than the inner parts, due to lower gas densities and a shallower local 
potential well toward larger radii.\ In this framework, galaxies naturally develop significant negative abundance gradients and slightly positive stellar age gradients, 
with more massive galaxies having steeper gradient slopes than lower mass galaxies.\ In the modern $\Lambda$ cold dark matter ($\Lambda$CDM) paradigm 
of hierarchical structure formation, present-day galaxies have formed their central parts first, through either cold stream accretion or violent disk instabilities 
(e.g., \citealt{Dekel_2009}) or galaxy merging, followed by a gradual buildup of galaxy disks from the inside out (e.g., \citealt{Mo_1998,Taylor2017}), 
which gives rise to negative radial gradients of both metallicities and stellar ages.\ Galaxy mergers are expected to play an important role (especially) at early cosmic 
epochs in the $\Lambda$CDM framework.\ Gas-poor mergers of comparable-mass galaxies (i.e., major mergers) can flatten preexisting stellar population gradients 
(\citealt{Kobayashi2004, DiMatteo2009}), while gas-rich major mergers may preserve or regenerate negative gradients (\citealt{Bekki1999,Hopkins2009}).\

With the advent of large integral field unit (IFU) spectroscopic surveys in the past decade, such as the Calar Alto Large Integral Field Area (CALIFA; \citealt{Sanchez2012}), Sydney Australian Astronomical Observatory Multi-object Integral Field Spectrograph (SAMI; \citealt{Bryant2015}), and Mapping Nearby Galaxies at the Apache Point Observatory (MaNGA; \citealt{Bundy2015}), it has become possible to obtain the spatial distribution of luminosity-weighted stellar ages and chemical abundances of large samples of galaxies in the local universe, which enables a direct investigation of the connection between stellar population gradients and other galaxy properties and thus a relatively straightforward test of galaxy formation models in general.

On one hand, the stellar population gradients are expected to be affected by the internal properties of galaxies.\ 
Studies of the radial profiles or gradients of the stellar ages and metallicities for over a hundred CALIFA galaxies \citep[e.g.,][]{Perez2013,GonzalezDelgado2014} revealed that the assembly of stellar mass and the cessation of star formation proceed from the galactic centers to the outskirts for massive galaxies, in line with the so-called ``inside-out'' growing mode, and there seems to be a negative correlation between stellar population gradients and stellar mass, which may naturally connect to the ``outside-in'' mode found for many nearby dwarf galaxies \citep{Zhang2012}.\
These findings agree well with previous studies of the radial stellar gradients based on multiwavelength broadband photometry (e.g., \citealt{Tortora2010,Gonzalez-Perez2011,Pan2015a,Pan2016}).\ 
Recent studies found more complicated dependences of stellar population gradients on stellar mass or other properties based on larger samples from the CALIFA survey (e.g., \citealt{Benito2017,GonzalezDelgado2017}) and the MaNGA survey (e.g., \citealt{Ibarra-Medel2016,Goddard2017,Ellison2018,Wang2018a}).\ 
In particular, the radial gradients of metallicities were found to depend on morphologies (e.g., Hubble type) rather than stellar mass in \citet{GonzalezDelgado2015}.\ 
\citet{Woo2019} found that the radial gradients of stellar age, specific star formation rates (sSFR), and abundance (O/H) for isolated galaxies rely on the galaxy's position on the diagram of stellar mass surface density within 1 kpc ($\Sc$) and stellar mass.\
Based on a study of 62 spiral galaxies, \cite{Sanchez-Blazquez2014} claimed that there is no dependence of stellar population radial gradients on stellar mass or morphological types when gradients are normalized to the effective radius of galaxy disks.\

On the other hand, the environment of galaxies might also play a nonnegligible role in regulating the radial stellar population gradients.\
 \citet{Schaefer2017} found that the SFR gradients are steeper in higher local galaxy density based on $201$ SFGs from the SAMI survey.\
However, the radial gradients of stellar age and metallicity have also been found to be independent of environmental parameters (e.g., the large-scale structure type and the local density in \citealt{Zheng2017}; the tidal strength parameter and the central--satellite division in \citealt{Goddard2017a}).\

Above all, it is not clear whether and how different galaxy properties (either internal or external) affect the radial distribution of the stellar population in nearby galaxies.\ 
In this work, taking advantage of IFU observations of a large sample of nearby galaxies from the most recent data release of SDSS-IV MaNGA survey, we attempt to 
explore the connection of stellar population radial gradients with other galaxy properties such as stellar mass, star formation, morphologies, and environments.\ Unlike 
most previous surveys that are either flux limited or volume limited, the MaNGA survey is designed to have a more or less uniform coverage in stellar mass.\ A series of 
recent studies have used the stellar population radial profiles of MaNGA galaxies to explore the general topic of galaxy quenching \citep[e.g.,][]{Li2015,Goddard2017,
Belfiore2018,Wang2018a}.\ The current paper focuses on a straightforward question: {\it What physical properties are closely connected to the observed stellar 
population gradients of nearby galaxies?}

The outline of this paper is as follows.\ In Section \ref{sec:data}, we describe our sample selection, determination of the radial gradients of D4000 indices, 
luminosity-weighted stellar ages, and luminosity-weighted stellar metallicities, as well as various morphological and environmental properties to be explored in this work.\ The results 
are presented in Section \ref{sec:results}, and the summary and discussion of our primary results follow in Section \ref{sec:discuss}.\

\section{Data}
\label{sec:data}
\subsection{Sample and Selection} 
\label{sec:sample}

As one of the three core programs in the fourth-generation Sloan Digital Sky Survey (SDSS-IV; \citealt{Albareti2017,Blanton2017}), MaNGA \citep{Bundy2015,Yan2016a,Yan2016} is an ongoing integral field spectroscopic (IFS) survey for obtaining two-dimensional spectral mapping of $\sim$ 10,000 nearby galaxies ($0.01 < z < 0.15$).\
MaNGA covers the wavelength range between $3600$ and $10300 \rm{\AA}$ with a spectral resolution of R $\sim$ 2000 \citep{Drory2015}.\ 
The sizes of the IFUs vary from 19 to 127 fibers, covering $12''-32''$ on the sky.\ The effective spatial resolution is 2\farcs5 \citep{Law2015,Law2016}. 

\begin{figure*}
   \includegraphics[width=0.95\textwidth]{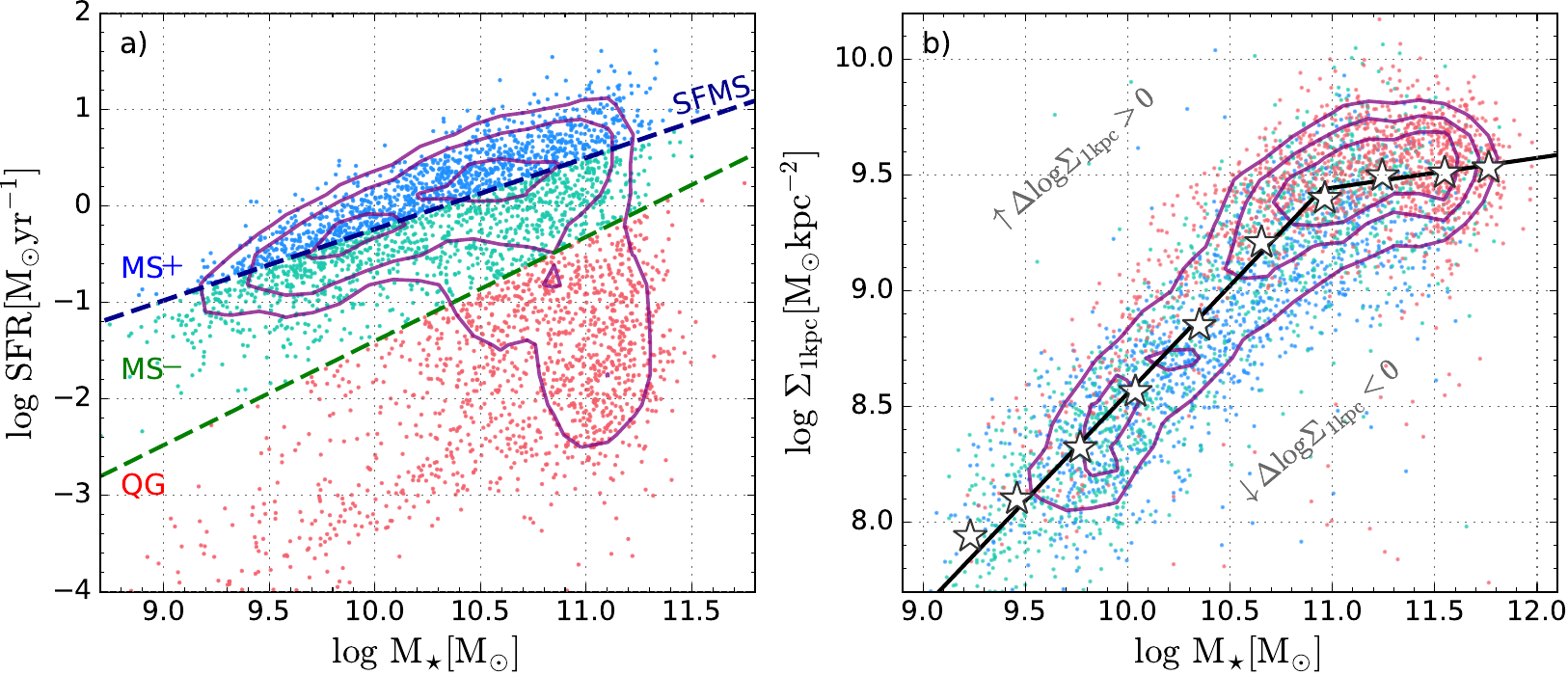}
   \begin{center}
     \caption{a) Distribution of our sample galaxies on the ${\rm SFR}$--$\Ms$ plane.\ 
     Star-forming galaxies and quiescent galaxies are divided by the green dashed line ($\log {\rm SFR} = 1.08 \log\Ms - 12.2$), and the star-forming galaxies are further divided into those above (MS$+$) and below (MS$-$) the SFMS relation (blue dashed line; $\log {\rm SFR} = 0.74 \log\Ms - 7.64$) adapted from \cite{Cano_2019}.\
     The purple contours enclose the central $25\%-50\%-75\%$ quantities of galaxies.\ %
     b) Distribution of our sample on the $\Sc$--$\Ms$ plane.\ The color scheme is as in the left panel. 
     The solid black bent line is the best-fit $\Sc$--$\Ms$ relation for our full sample ($\log\Sc = 0.93(\log\Ms-10.95) + 9.44$ for $\log\Ms \le 10.95$, $\log\Sc = 0.13(\log\Ms-10.95) + 9.44$ for $\log\Ms>10.95$).\
     The white star symbols mark the median $\log$ $\Sc$ of galaxies falling into individual 0.3 dex stellar mass bins.
     } 
        \label{Sample}
   \end{center}
  \end{figure*}

In this work, we use the reduced data from the MaNGA Pipe3D Value Added Catalog (VAC; \citealt{Sanchez2018}) available in the SDSS DR15 (\citealt{Aguado2019}).\
The SDSS DR15 includes 4656 MaNGA galaxies.\
The Pipe3D provides spectroscopic redshift, star formation rate (SFR) derived from the $\ha$ luminosities, and stellar mass for each galaxy.\ 
Because the Pipe3D SFRs and stellar masses are consistently estimated based on the MaNGA data, we use their ratios to quantify the specific sSFR for our galaxies.\ In addition, the NASA Sloan Atlas catalog (NSA; \citealt{Blanton2011}) provides auxiliary galaxy properties, such as the galaxy effective radius $(\Reff)$, axis ratio $(b/a)$, position angle, S\'{e}rsic index $(n)$ measured in the $r$ band, and stellar masses for SDSS galaxies.\ The NSA stellar masses are determined based on the SDSS imaging data and are expected to be a better representation of the total stellar masses than those based on the MaNGA data.\ Throughout this paper, we use the NSA stellar masses $\Ms$ except when exploring the mass--SFR relations (Figure \ref{Sample}a), where the Pipe3D stellar masses (and SFR) are used.\ Because the Pipe3D stellar parameters used in this work are based on the \citet{Salpeter1955} initial mass function (IMF), we adjust the NSA stellar masses, which were originally based on the \cite{Chabrier2003} IMF, to be consistent with the \citet{Salpeter1955} IMF.\

Following the quality control catalog included in the Pipe3D VAC, 84 of the 4656 galaxies have either wrong redshift, poor signal-to-noise ratio (S/N), or other warning issues and are thus excluded from our sample.\ These 84 galaxies are excluded from our sample.\
In addition, 532 galaxies that were observed with the 19-fiber IFUs are also excluded from our final sample because the radial gradients estimated using the low spatial-resolution IFUs may be subject to relatively large biases \citep[e.g.,][]{Ibarra-Medel2018}.\  
We also exclude $140$ galaxies that have minor-to-major axis ratios $(b/a)$ less than $0.3$. 
We further restrict our analysis in this paper to galaxies with $0.01< z < 0.15$, $10^{9}~\Msun \leq \Ms \leq 10^{12}~\Msun$, and $10^{-5}~\Msunyr < {\rm SFR} < 10^{4}~\Msunyr$ and abandon galaxies without calculations of radial stellar population gradients (Section \ref{sec:meth}), which leaves us with 3654 galaxies in total.\ 
Both the ``Primary" and ``Secondary" MaNGA samples are used in this work.\

Figure \ref{Sample}a) shows the stellar mass--SFR distribution of our sample galaxies.\
We divide our sample into galaxies above (MS$+$; 1132 galaxies) and below (MS$-$; 1222 galaxies) the galaxy star formation main sequence (SFMS) relation, and quiescent galaxies (QGs; 1300 galaxies) by using the lines of demarcation from \citet{Cano_2019}.\

\subsection{Radial stellar population gradients}
\label{sec:meth}

Two-dimensional maps of the 4000\AA~break (D4000) and stellar population parameters, including the surface density of stellar mass, linear values of the luminosity-weighted stellar age ($\SA$), and logarithmic values of the luminosity-weighted stellar metallicity ($\SZ$), are retrieved from Pipe3D pipeline.\ We choose to use the D4000 index as a representative age-sensitive observable, because this broad feature can be measured with relatively 
small uncertainties ($<$ 10\%) compared to other commonly used absorption features (e.g.,\ H$\delta$ line) at spectral S/Ns 
down to $\gtrsim$ 1/\AA~(\citealt{Cardiel1998}).\ This practical advantage is particularly relevant for a comparative analysis of galaxies or regions with different luminosities or surface brightnesses, as is presented in this work.\ 

The model-dependent stellar population parameters $\SA$ and $\SZ$ from Pipe3D are derived through stellar population synthesis modeling of the MaNGA spectra, as described in \citet{Sanchez2016a,Sanchez2016}.\ Pipe3D also produces mass-weighted stellar ages and metallicities.\ 
Generally speaking, stellar population parameters estimated from integrated spectra of unresolved (sub)galactic regions are biased toward younger stellar populations, and the constraints on older stellar populations are unavoidably subject to larger uncertainties.\ On the other hand, the stellar mass contribution of relatively old stellar populations in local galaxies usually dominates over that of young stellar populations.\ Therefore, the very poor constraints on star formation history (SFH) at ancient times means that the mass-weighted stellar parameters derived from integrated spectra are subject to substantially larger uncertainties than the luminosity-weighted ones, which may frustrate any attempt to obtain useful estimates of mass-weighted parameters for a comparative study as presented in this work.\ Throughout this paper, our discussion of stellar ages and metallicities is focused on the luminosity-weighted ones, but we also present the relevant results based on mass-weighted ones in the Appendix (Figures \ref{RFimportance_MW}, \ref{Grad_MW}, and \ref{Grad_Mass_MW}).\

We derive the radial gradient slopes of the D4000, $\SA$, and $\SZ$ of each galaxy by performing linear least-squares fitting to the median radial profiles.\
Specifically, we first calculate the deprojected galactocentric distance of each spaxel by using the observed axis ratio of its host galaxy.\
Then, we obtain radial profiles of D4000, $\SA$ and $\SZ$ by calculating the median D4000, $\SA$ and $\SZ$ of spaxels falling within the 2\farcs5 wide concentric elliptical annuli.\ The radial gradient slopes are derived via linear least squares fitting to the median radial profiles, defined as 
\begin{equation}
\nabla{\rm{D4000}} = \frac{d{\rm D4000}}{d(R/R_{\rm e})},
\label{eq-gradD4000}
\end{equation}
\begin{equation}
\nabla{\tau_{\rm{L}}} = \frac{d\tau_{\rm L}\rm{[Gyr]}}{d(R/R_{\rm e})},
\label{eq-gradAge}
\end{equation}
\begin{equation}
\nabla{{[Z/{\rm H}]_{\rm{L}}}} = \frac{d[Z/{\rm H}]_{\rm{L}}}{d(R/R_{\rm e})},
\label{eq-gradZH}
\end{equation}
where $R/R_{\rm e}$ is the radius in units of the effective radius.\
A negative $\GD$ represents a larger D4000 in the galactic center compared to its outskirts, corresponding to the inside-out scenario, while a positive $\GD$ is consistent with the outside-in scenario.\
We note that only spaxels with S/N $\geq$ 10/\AA~are used for calculating the medians, and only radial bins containing $\geq$ 5 data points with 
S/N $\geq$ 10/\AA~are used in the gradient determination.\ Although the MaNGA ``Secondary" sample observations extend to $\sim$ 2.5$R_{e}$, 
we only analyze their radial profiles out to $\sim$ 1.5$R_{e}$ in order to be consistent with the ``Primary" sample.

It is worth noting that we choose to quantify the gradients of ages on linear scales instead of the logarithmic scales that were adopted in many previous studies.\ The same linear age gradient corresponds to flatter logarithmic age gradients at older ages than at younger ages, which may give one a misleading impression that stellar age gradients flatten over time.\ Indeed, we found that the logarithmic age gradients become significantly flatter for galaxies with lower sSFR (not shown here), but the linear age gradients barely show such a trend (Figure \ref{Grad}).

\subsection{Morphological Properties}
\label{sec:struc}

In order to quantify the morphological dependence of stellar population radial profiles of our galaxies, we consider four morphological parameters, 
i.e.,\ T-Type, stellar mass surface density averaged within the central 1 kpc ($\Sc$), central velocity dispersion ($\sigma_{\rm{cen}}$), and S\'{e}rsic index $n$ 
\citep{Sersic1968}.\ The T-Type values are retrieved from the MaNGA PyMorph Photometric and Deep Learning Morphological Catalogs (\citealt{Fischer2019}). 
$\Sc$ is calculated by summing up the stellar mass map within the central $1\kpc$ of each galaxy.\ $\sigma_{\rm{cen}}$ is retrieved from the Pipe3D VAC (\citealt{Sanchez2018}), 
and the S\'{e}rsic index $n$ is retrieved from the NSA catalog (\citealt{Blanton2011}).

The T-Type values of our sample galaxies are quantified as $T = -4.6\times P({\rm Ell})-2.4\times P({\rm S0}) +2.5\times P({\rm Sab})+6.1\times P({\rm Scd})$ \citep{Meert2015}, where $P$ is the probability of the morphological class indicated in parentheses \citep{Huertas2011}.\ This T-Type definition has been 
extensively used to quantify the visual morphologies of galaxies \citep[e.g.,][]{Nair2010,Willett2013,Dominguez2018}.\ According to this definition,  
elliptical and S0 galaxies (ETG) have T-Types $\leq$ 0, while late-type disk galaxies (LTG) have T-Types $>$ 0.\ 

$\Sc$ has been widely used to quantify the central stellar mass surface densities (\citealt{Fang2013,Wang_2018,Woo2019}).\ There is a well-established 
correlation between $\Sc$ and $\Ms$ (e.g.,\ Figure \ref{Sample}b).\ We perform piecewise linear least-squares fitting to $\log\Sc$ as a function of $\log\Ms$ 
for our galaxies.\ The best-fit broken linear relation is $\log\Sc = 0.93(\log\Ms-10.95) + 9.44$ for $\log\Ms \le 10.95$, $\log\Sc = 0.13(\log\Ms-10.95) + 9.44$ for $\log\Ms>10.95$.\
In what follows, we use the vertical offset of $\log\Sc$ ($\DSc$) with respect to this best-fit $\log\Sc$--$\log\Ms$ relation to quantify the relative excess or deficit 
of the central stellar mass of each galaxy for its $\Ms$.\ We note that, at given stellar masses, the $\DSc$ difference is more or less equivalent to the logarithmic difference of the bulge-to-total mass ratios ($B/T$).

\subsection{Environmental Parameters}
\label{sec:envir}
In order to explore the environmental dependence of stellar population radial profiles, we separate our sample into central and satellite galaxies by using the halo-based 
group catalog of SDSS galaxies derived by \citet{Lim2017k}, and then use the projected local galaxy overdensity within a projected distance to the fifth nearest neighbor 
to quantify the local environment of each galaxy.\ The \citet{Lim2017k} group catalog improves upon that of \citet{Yang2005f, Yang2007} and \citet{Lu2016}, with a more 
uniform group halo assignment over a wider range of halo masses $M_{\rm{halo}}$.\ The local overdensities are quantified as $\log (1 + \delta) = \log \left(1 + (\rho_{i} - \rho_{m})/\rho_{m}\right)$ \citep{Etherington2015}, where $\rho_{i}$ is the local galaxy number density and $\rho_{m}$ is the mean galaxy number density.\ 
Among our sample galaxies, 
2569 have $\log (1 + \delta)$ measurements available in the Galaxy Environment for MaNGA Value Added Catalog (GEMA-VAC, Argudo-Fernádez et al. in prep.), and among these 2569 galaxies, 2543 have available central--satellite division (1699 centrals and 844 satellites).

\section{Results}
\label{sec:results}

This section is devoted to an exploration of the connection of the radial profiles of D4000, $\SA$ and $\SZ$ with the various galaxy properties introduced above.\ 
We first present an evaluation of the relative importance of different galaxy properties in predicting $\GD$, $\GA$, and $\GZ$, as well as the average D4000, 
$\SA$, and $\SZ$ within 0.5 effective radius (i.e.,\ $\SDC$, $\SAC$ and $\SZC$) and around one effective radius (i.e.,\ $\SDRE$, $\SARE$ and $\SZRE$), 
of our sample galaxies using a Random Forests (RF; \citealt{Breiman2001_RF}) algorithm, and then present the dependence of stellar population profiles 
on the most relevant galaxy properties separately.

\subsection{The Most Predictive Properties for Stellar Population Profiles}
\label{sec:randomforest}

\begin{figure*}[ht!]
   \begin{center}
   \includegraphics[width=1.0\textwidth]{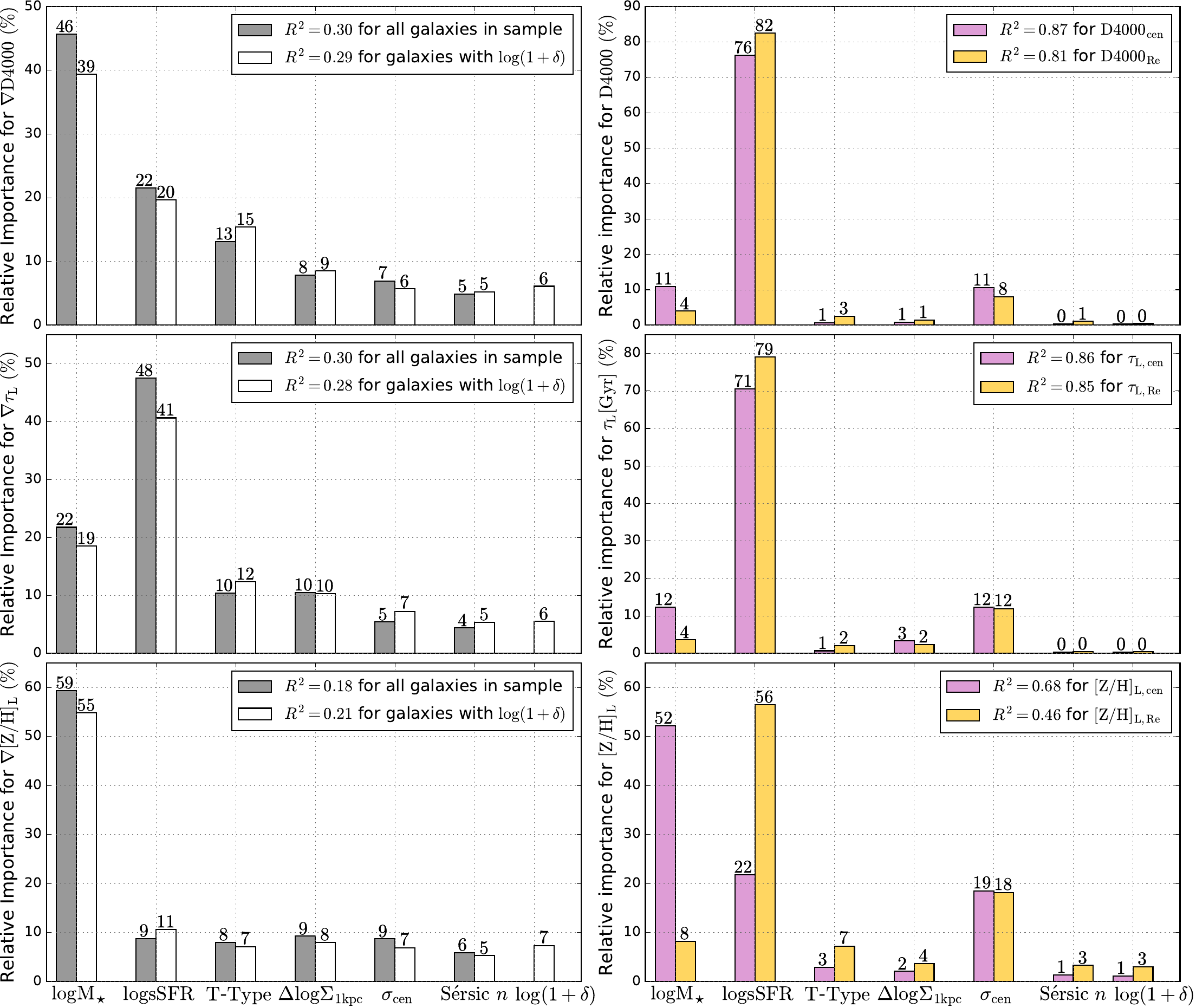}
      \caption{
       Left column: evaluation of the relative importance of different galaxy properties in predicting the radial gradients $\GD$ (top), $\GA$ (middle), and $\GZ$ (bottom) of our 
      galaxies by using the RF estimator.\ 
      The gray bars in each panel represent results for the whole sample, and the white bars represent results for the subsample with $\log (1+\delta)$ measurements.\ 
      Right column: evaluation of the relative importance of different galaxy properties in predicting D4000, $\SA$, and $\SZ$ 
      at the galaxy center (purple bars) and one effective radius (orange bars) of our sample galaxies with $\log(1+\delta)$ measurements.\
      The coefficient of determination $R^2$ returned from each RF run is indicated in each panel.\
      See the text for more details.
      } 
      \label{RFimportance}
   \end{center}
\end{figure*} 

We adopt the RF estimator implemented in the Python package {\sc scikit-learn} (\citealt{Pedregosa_2011}) to evaluate the relative importance of different galaxy 
properties in predicting $\GD$, $\GA$, and $\GZ$, as well as $\SDC$, $\SAC$, $\SZC$, $\SDRE$, $\SARE$, and $\SZRE$.\
The RF algorithm is an ensemble method commonly used in astrophysics not only for classification (e.g., \citealt{Dubath2011,Richards2011}), but also in regression (e.g., \citealt{Miller_2015}), where the prediction can be for a real-valued response variable.\ More details of the decision tree regression can be found in \citet{Hastie2009} and \citet{Kuhn2013}.
Basically, the RF estimator fits a number of decision tree classifiers on various subsamples randomly drawn with replacement from our galaxy sample and uses the averaging of individual decision trees to improve the predictive accuracy.\ Every decision tree in the ensemble is trained by using a random selection of features (i.e., the galaxy properties we are interested in) to split galaxy samples belonging to the tree.\ The more often a feature is used in splitting galaxies of a tree, the more important that feature is, and the final relative importance of the different galaxy properties in predicting our stellar population parameters of interest is simply the average of feature importances of individual trees in the ensemble.\

Because a fraction of our galaxies do not have $\log(1+\delta)$ measurements, we run the RF estimator separately on the full sample to evaluate the relative importance 
of $\Ms$, sSFR, T-Type, $\DSc$, $\sigma_{\rm{cen}}$, and S\'ersic index $n$, and on the subsample of 2569 galaxies with $\log(1+\delta)$ measurements for the relative 
importance of $\Ms$, sSFR, T-Type, $\DSc$, $\sigma_{\rm{cen}}$, $n$ and $\log(1+\delta)$.\ The most important parameters for setting up the RF estimator are the number of decision trees (n\_estimators) and the minimum number/fraction of samples for node splitting (min\_samples\_split).\ 
The minimum number/fraction of samples at a splitted leaf node (min\_samples\_leaf) is set to be min\_samples\_split/2.\
After extensive testing, we find that the final results do not change with n\_estimators as long as n\_estimators $\geq$ 100.\ 
The exact values of the feature importance vary with min\_samples\_split, but we note that the overall ranking of the relative importances of the top features (or galaxy properties) do not change qualitatively for our galaxy sample, as long as a reasonably small or large min\_samples\_split is used (e.g., min\_samples\_split = 0.1\% $-$ 10\% of the number of our galaxies).\ Lastly, we mention that the two criteria for measuring the quality of tree splitting ``MSE'' and ``MAE'' supported in the RF estimator give nearly the same results for our samples, and we choose to use the ``MSE'' criterion for presenting our results.\

Figure \ref{RFimportance} presents the RF ranking results by using n\_estimator = 1000 and min\_samples\_split = 1\%.\ The left column shows the results for radial gradients, while the right column shows the results for average stellar population parameters at the galaxy center and near one effective radius, respectively.\
For the sake of clarity, only subsamples with local overdensity measurements are considered for the average stellar population parameters.\ 
The coefficient of determination $R^2$ for each RF run, which quantifies the fraction of variance of the data that is ``explained'' by the model, is indicated in Figure \ref{RFimportance}.\
$R^2$ normally ranges from 0.0 to 1.0 (best), but may go negative when the model is arbitrarily worse.\
The relative feature importance for a given RF run is normalized such that the sum for all the explored properties is equal to one.\ 

\subsubsection{Ranking of galaxy properties for radial gradients}

For radial gradients, it is obvious that $\Ms$ is the most predictive property for $\GD$ (top-left panel of Figure \ref{RFimportance}), and the next two most predictive properties for $\GD$ are, in order of decreasing importance, sSFR and T-Type.\ The first three most predictive properties combined account for $\sim$ 80\% of the total feature importances.\
It is noteworthy that local galaxy overdensities have a very weak correlation with $\GD$.\
We also explored the importance of being central or satellite galaxies (not shown here) and found that the central-satellite division has almost zero predictive power for $\GD$.\

D4000 varies with both stellar ages and metallicities.\ 
The middle and bottom panels of the left column of Figure \ref{RFimportance} show the RF ranking results for $\GA$ and $\GZ$, respectively.\ sSFR and, to a lesser degree, 
$\Ms$ are the two most predictive properties for $\GA$, while $\Ms$ is the single most predictive property for $\GZ$.\ Therefore, the apparent dependence of $\GD$ on $\Ms$ 
is primarily a metallicity effect, while the apparent dependence of $\GD$ on sSFR is primarily an age effect.\ 

\subsubsection{Ranking of galaxy properties for stellar populations at center and one effective radius}
As is shown in the right column of Figure \ref{RFimportance}, the sSFR is the single most predictive property for D4000 and $\SA$ at both the galaxy center and one effective radius 
for our sample.\ As the second most predictive property for local D4000 and $\SA$, the central velocity dispersion $\sigma_{\rm{cen}}$ has a significantly smaller 
predictive power than sSFR.\ For central $\SZ$, $\Ms$ is the most predictive property.\ However, at one effective radius, sSFR is the single most predictive property for $\SZ$, similar to D4000 and $\SA$.

\subsection{Dependence of Stellar Population Gradients on Stellar Mass and sSFR}
\label{sec:mass}

\begin{figure*}[ht!]
   \begin{center}
   \includegraphics[width=1.0\textwidth]{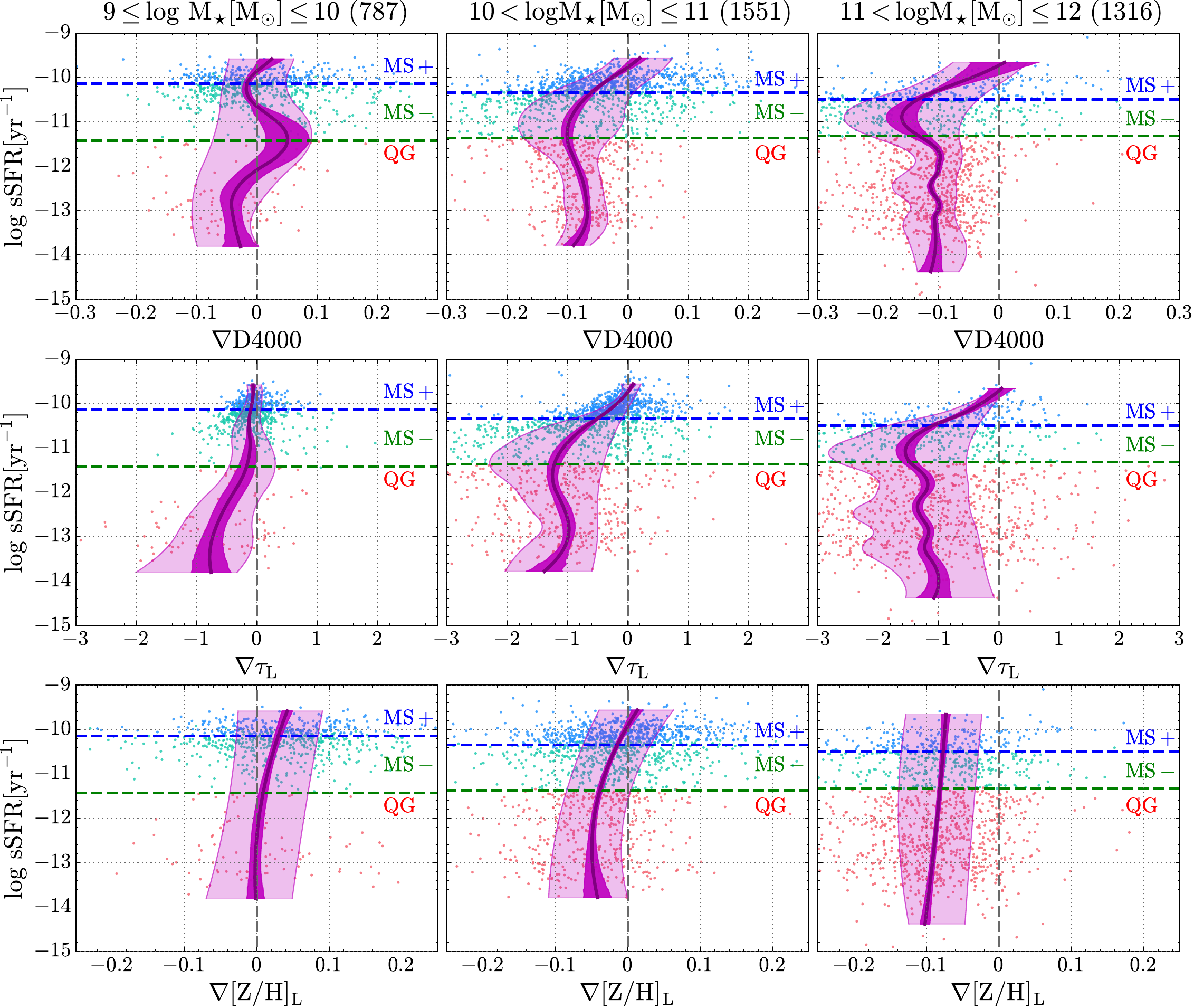}
      \caption{Global sSFR is plotted against stellar population gradients (from top to bottom: $\GD$, $\GA$, and $\GZ$) for galaxies in different $\log\rm{M_\star}$ bins, as indicated at the top of each column.\ 
      In each panel, the thick solid curve represents the median trend, and the dark shaded region marks the uncertainties of the median trend.\ The light shaded regions are bounded by the smoothed curves of the $25\%-75\%$ quantiles of the galaxy distribution.\ Both the median and $25\%-75\%$ quantiles are determined with the nonlinear spline quantile regression method.\ The uncertainties of the median trend are determined based on the random resampling of the original sample with replacement 1000 times.\
      The two dashed horizontal lines divide the sample into MS$+$ (blue dots), MS$-$ (green dots), and QG (red dots) subsamples.\ 
      The number of galaxies in each stellar mass bin is given in parentheses at the top of the corresponding column.
      }
      \label{Grad}
   \end{center}
\end{figure*} 

\begin{figure*}[ht!]
   \begin{center}
   \includegraphics[width=1.0\textwidth]{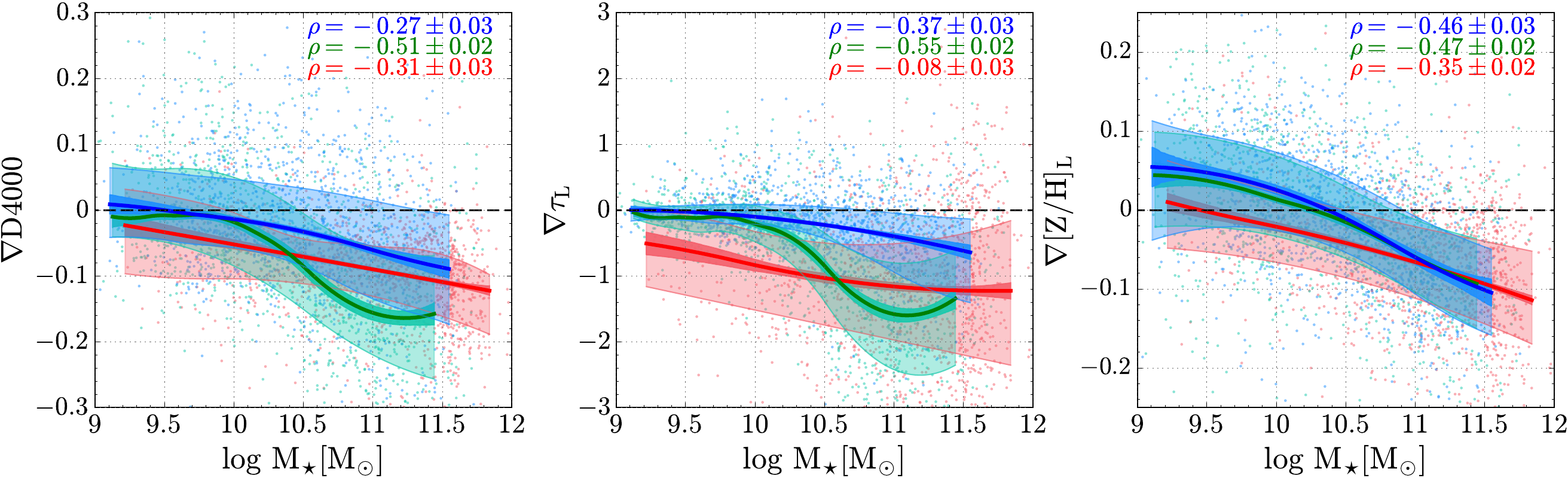}
      \caption{
      Stellar population gradients (from left to right: $\GD$, $\GA$, and $\GZ$) as a function of stellar masses for MS$+$ (blue), MS$-$ (green), and QG (red).\ 
      As in Figure \ref{Grad}, the thick solid curves in each panel represent the median trend for different subsamples, and the dark shaded regions mark the uncertainties of the median trend.\ The light shaded regions mark the smoothed $25\%-75\%$ quantiles of the galaxy distribution.\ Both the median and $25\%-75\%$ quantiles are determined with the nonlinear spline quantile regression method.\ The Spearman's ranking correlation coefficients $\rho$ for different galaxy samples and their standard deviations determined based on 1000 Monte Carlo realizations are indicated in each panel.\ 
      }
      \label{Grad_Mass}
   \end{center}
\end{figure*} 

The whole sample is split into three stellar mass bins: $9\leq\log\Ms\leq10$ (low-mass galaxies), $10<\log\Ms\leq11$ (intermediate-mass galaxies),  
and $11<\log\Ms\leq12$ (high-mass galaxies).\ Figure \ref{Grad} shows the variation of $\GD$ (top row), $\GA$ (middle row), and $\GZ$ (bottom row) 
with integrated sSFR for galaxies in the three stellar mass bins (from the first to the third columns).\ 
In each panel, MS$+$, MS$-$, and QGs are plotted with blue, green, and red symbols, respectively.\ 
Smoothing curves of the $25\%-50\%$(median)$-75\%$ quantities for $\GD$, $\GA$, and $\GZ$ along the sSFR axis are calculated by using the Comprehensive R Archive Network (CRAN) package Constrained B-Splines (COBS; \citealt{COBS_2007,COBS_2020}), which gives qualitatively constrained smoothing spline curves.\ 
To determine the uncertainties of the median curves, we randomly resample with replacement the original sample for 1000 times and perform the spline quantile regression to each realization.\

For the intermediate- and high-mass bins, the median $\GD$ gradually varies from nearly zero to increasingly more negative values as sSFR decreases, reaches the 
minimum at $\sim 0.5-1$ dex below the SFMS, and finally ``saturates'' at a more or less constant negative value as sSFR falls $\gtrsim$ 1 dex below the SFMS, corresponding 
roughly to the regime occupied by QGs.\ In contrast to the higher mass bins, the median $\GD$ of the low-mass galaxies stays close to zero until sSFR falls $>$ 2 dex below the 
SFMS.\ In addition, the median $\GD$ values at a given sSFR are smaller (i.e.,\ more negative) for more massive galaxies.\ The median $\GA$ exhibits a similar variation with 
sSFR at given $\Ms$, in line with the RF ranking results that the sSFR dependence of $\GD$ is primarily driven by stellar age effect (Section \ref{sec:randomforest}).\ 
The median $\GZ$ becomes more negative as sSFR decreases for the intermediate-mass galaxies but stays more or less constant for the low- and high-mass galaxies.

Figure \ref{Grad_Mass} further illustrates a continuous mass-dependent variation of the median $\GD$, $\GA$, and $\GZ$ for the MS$+$, MS$-$, and QG subsamples.\
   Smoothing curves depicting the $25\%-50\%$(median)$-75\%$ quantities of the galaxy distributions as well as the uncertainties of the medians for different subsamples are calculated in the same way as in Figure \ref{Grad}.\
The mass-dependent variations for star-forming galaxies generally follow much shallower slopes at $\log\Ms$ $\lesssim 10.0-10.5$, above which the MS$-$ subsample exhibits significantly steeper slopes than the MS$+$ subsample for $\GD$ and $\GA$.\ It is noteworthy that while the median $\GZ$ monotonically decreases with $\Ms$ across the whole mass range, the median $\GA$ of MS$-$ galaxies reaches the most negative values at $\log\Ms$ $\sim$ 10.8 and stays more or less constant at larger masses.\ 

We further use the Spearman's rank correlation coefficients $\rho$ to measure the correlation between stellar population gradients and stellar mass for each subsample.\ 
We also determine the standard deviation of $\rho$ based on random resampling of the original samples with replacement.\ As indicated in Figure \ref{Grad_Mass}, there exist  
significant correlations ($\rho > 0.2$) between the explored radial gradients and stellar masses, except for the $\GA$ of the QG subsample.\

\subsection{Dependence of Stellar Population Profiles on Galaxy Morphologies}
\label{sec:mor}

\begin{figure*}[ht!]
   \begin{center}
    \includegraphics[width=0.85\textwidth]{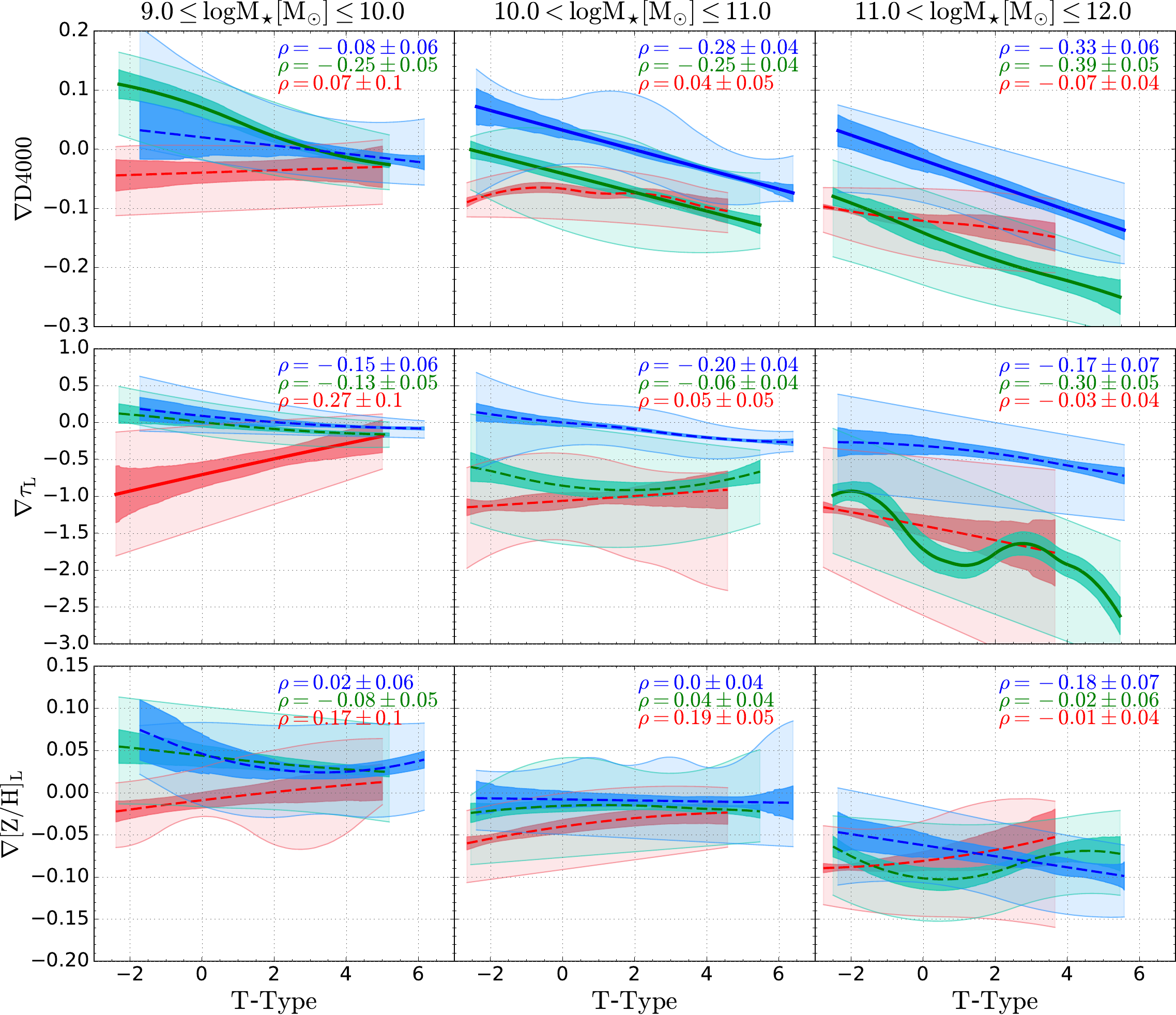}
       \caption{Dependence of stellar population gradients on T-Types.\ Results for $\GD$, $\GA$, and $\GZ$ are shown respectively in the {\it top}, {\it middle}, and {\it bottom} rows.\ 
       Galaxies that fall in the low ($9\leq\log\Ms\leq10$), intermediate ($10<\log\Ms\leq11$) and high ($11<\log\Ms\leq12$) stellar mass ranges are plotted, respectively, in the {\it left}, {\it middle}, and {\it right} columns.\ 
       In each panel, MS$+$, MS$-$, and QGs are plotted with blue, green and red colors, respectively.
       The $25\%-50\%$(median)$-75\%$ quantities of the subsample galaxies with $1\sigma$ uncertainties of the median smoothing curve and the Spearman correlation coefficients $\rho$ with uncertainties for the subsamples are calculated in the same way as Figure \ref{Grad_Mass}.\ 
       Note that the existence of a correlation is only confirmed for $\rho > 0.2$ (solid line), and those with insignificant correlations are plotted as dashed lines.\ 
       }
       \label{Grad_TTYPE}
   \end{center}
\end{figure*} 

\begin{figure*}[ht!]
   \begin{center}
    \includegraphics[width=0.85\textwidth]{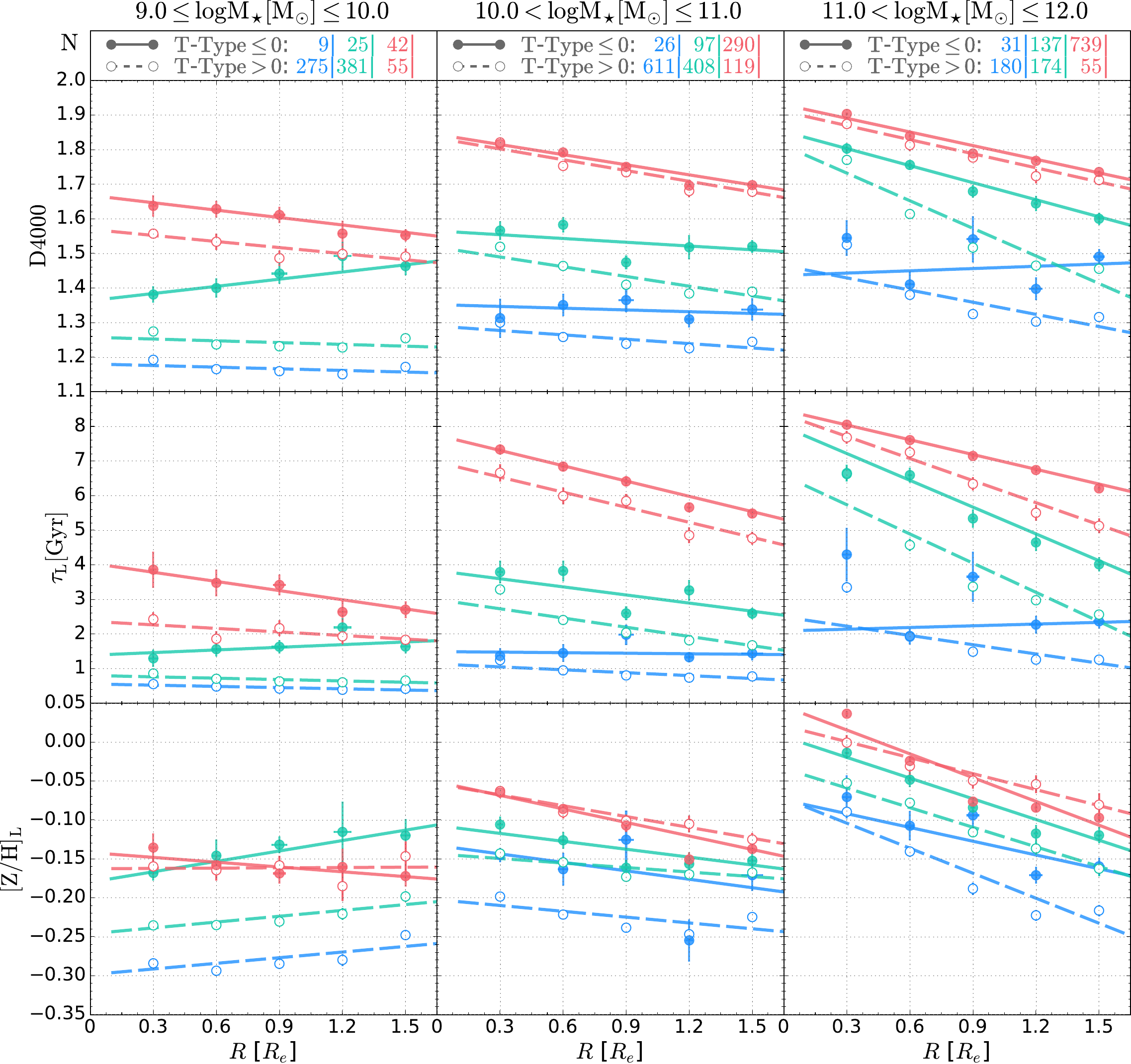}
       \caption{Median coadded radial profiles of D4000, $\SA$, and $\SZ$ for early-type galaxies (ETGs: T-Type $\le$ $0$; filled symbols) and late-type galaxies (LTGs: T-Type $>$ $0$; open symbols).\ The solid and dashed lines in each panel represent linear least-squares fitting to ETGs and LTGs, respectively.\ Division of the samples into different ranges of stellar mass and sSFR, and the color coding scheme are as in Figure \ref{Grad_TTYPE}.\ Median coadded radial profiles are plotted only when there are more than 15 data points in a given bin.\ The number of galaxies in different subsamples is also listed at the top of each column. \ 
       }
       \label{Profile_TTYPE}
   \end{center}
\end{figure*} 

\begin{figure*}[ht!]
   \begin{center}
    \includegraphics[width=0.85\textwidth]{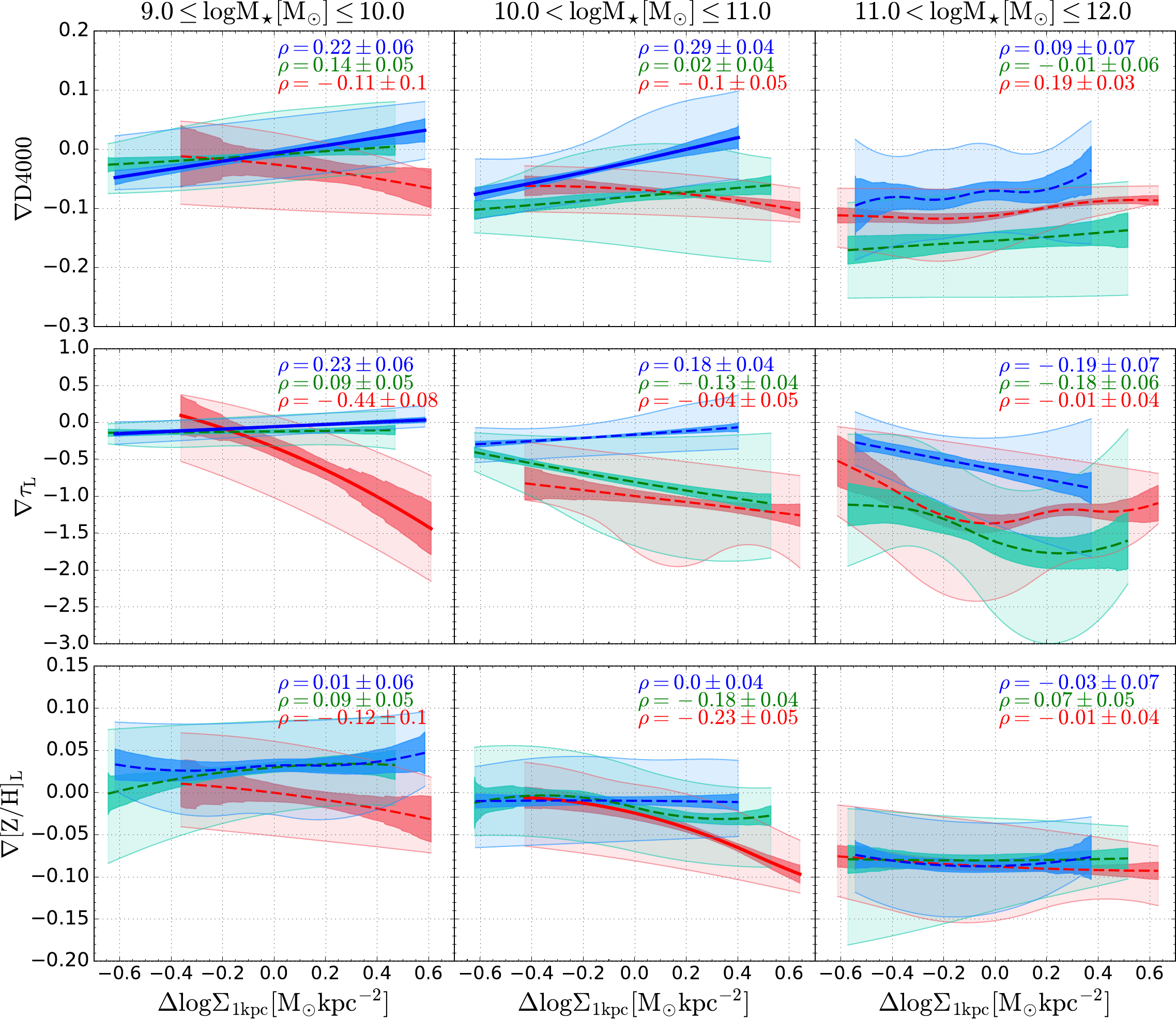}
       \caption{Same as Figure \ref{Grad_TTYPE}, but for the dependence of radial gradients of D4000, $\SA$, and $\SZ$ on $\DSc$.
       }
       \label{Grad_DS1}
   \end{center}
\end{figure*}

\begin{figure*}[ht!]
   \begin{center}
    \includegraphics[width=0.85\textwidth]{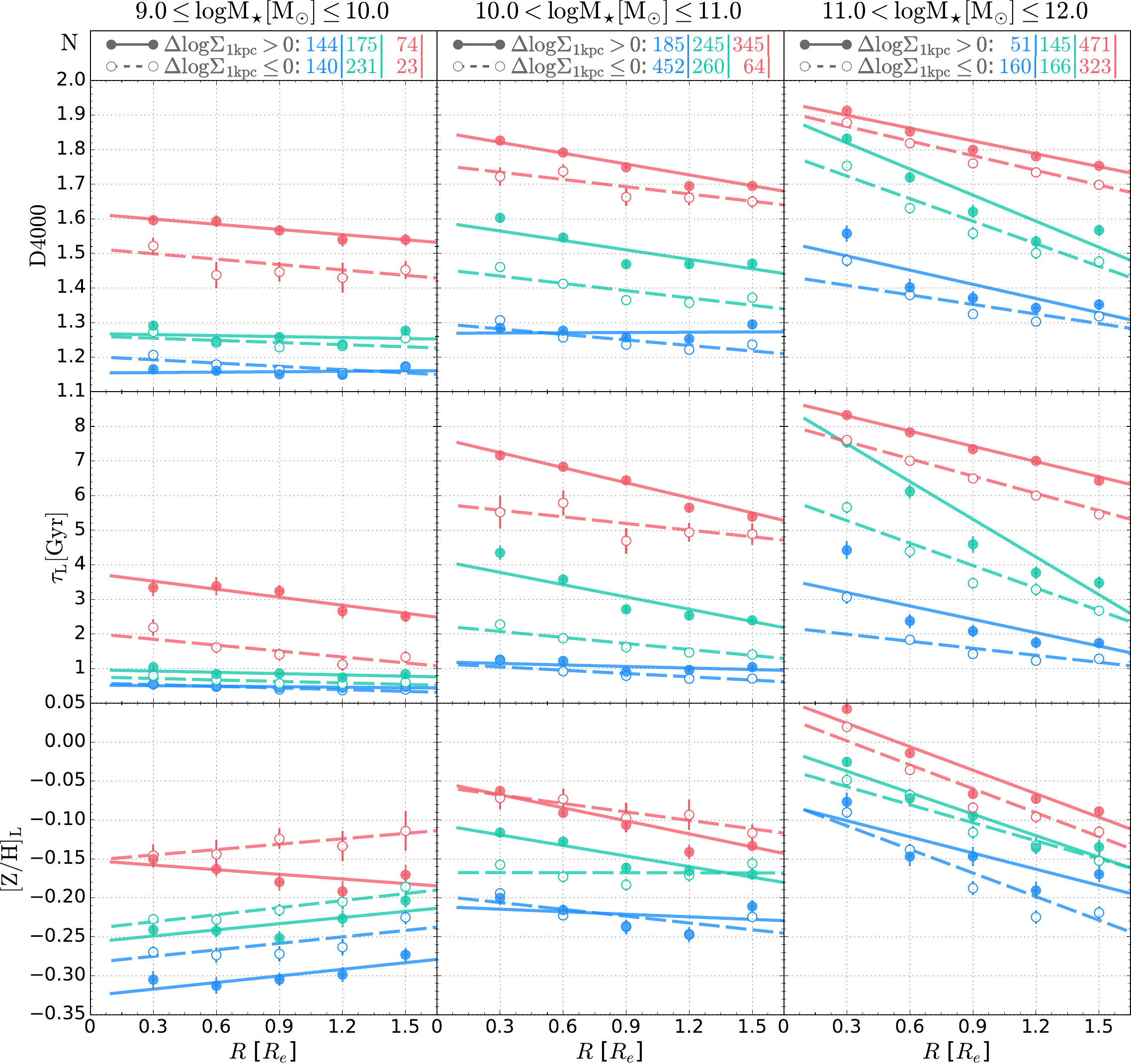}
       \caption{Same as Figure \ref{Profile_TTYPE}, but for the median coadded radial profiles of D4000, $\SA$, and $\SZ$ for galaxies with compact cores ($\DSc$ $>$ $0$; filled symbols) and diffuse cores ($\DSc$ $\le$ $0$; open symbols).
       }
       \label{Profile_DS1}
   \end{center}
\end{figure*} 
We have shown that stellar mass and sSFR are the two parameters that have the strongest correlation with stellar population gradients in previous sections.\ Here we explore the secondary dependence of stellar population gradients on morphological properties, as quantified by T-Type and $\DSc$, by controlling the stellar mass and sSFR.\
The dependence of the median $\GD$, $\GA$, and $\GZ$ on T-Type and $\DSc$ for MS$+$, MS$-$, and QGs in the three stellar mass bins defined in 
Section \ref{sec:mass} are presented in Figures \ref{Grad_TTYPE} and \ref{Grad_DS1}.\ The smoothing curves of the $25\%-50\%$(median)$-75\%$ quantities shown in Figures \ref{Grad_TTYPE} and \ref{Grad_DS1} are derived in the same way as in previous sections.\
Correlations with the Spearman's rank correlation coefficient $\rho > 0.2$ are regarded significant.\
In Figures \ref{Grad_TTYPE} and \ref{Grad_DS1}, median smoothing curves for subsamples with significant correlations are plotted as solid lines, whereas those with insignificant correlations are plotted as dashed lines. 

To illustrate the radial profile variations of D4000, $\SA$, and $\SZ$ with T-Type and $\DSc$, we divide our samples into separate bins of T-Type ($\le$ $0$ 
and $>$ $0$) and $\DSc$ ($>$ $0$ and $\le$ $0$), and show their respective median radial profiles in Figures \ref{Profile_TTYPE} and \ref{Profile_DS1}.\ 
In the following subsections, we present the dependence of the median radial profiles and gradients on T-Type and $\DSc$ separately. 

\subsubsection{Dependence on T-Type}
\label{subsec:TType}

As is shown in Figure \ref{Grad_TTYPE}, galaxies of higher stellar masses generally have more negative median $\GD$, $\GA$, and $\GZ$ at 
given T-Type and sSFR ranges.\ The median $\GD$ and $\GA$ are generally more negative at later T-Types for given mass and sSFR ranges.\ 
However, we point out that the overall negative correlation between $\GA$ and T-Types are not significant for all but the intermediate-mass MS+ and high-mass 
MS$-$ subsamples.\
There is no significant correlation between T-Types and $\GZ$ for any subsamples.
Nevertheless, the median D4000, $\SA$ and $\SZ$ of earlier-type (smaller T-Types) galaxies are generally larger than that of later-type (larger T-Types) galaxies across the probed radial ranges for any given mass and sSFR ranges (Figure \ref{Profile_TTYPE}).\ 

The overall more significant negative correlation between $\GD$ and T-Types than that between $\GA$ and T-Types may be qualitatively explained 
by the fact that the derivative of D4000 with respect to age decreases with age for a passively evolving stellar population.\ The same stellar age radial 
gradient results in a shallower D4000 radial gradient for earlier T-Types, which tend to have overall older ages.\ We can conclude here that the morphological 
type is a property that is connected with galaxy stellar populations largely in a radius-independent way.\

\subsubsection{Dependence on $\DSc$}
\label{subsec:DSc}
As with T-Types, galaxies of higher stellar masses generally have more negative median $\GD$, $\GA$ and $\GZ$ at given $\DSc$ and sSFR ranges.\
While the dependence of $\GD$ on $\DSc$ is generally not significant except for the low- and intermediate-mass MS$+$ subsamples, the median trend is generally in the opposite sense with that on T-Types (Figure \ref{Grad_DS1}), which probably reflects the expected negative correlation between T-Types and central mass concentrations in general.\ Particularly, the $\GD$ of the low- and intermediate-mass MS$+$ subsamples exhibit significant positive correlations with $\DSc$, and the median $\GD$ becomes positive at the high $\DSc$ end.\ Such positive dependence on $\DSc$ is also found for the median $\GA$.\ Therefore, the low- and intermediate-mass MS$+$ galaxies with higher $\DSc$ tend to have more centrally concentrated star formation activities.\ The correlation between $\GZ$ and $\DSc$ is generally very weak.\
Although the dependence of radial gradients on $\DSc$ is overall weak, galaxies with higher $\DSc$ have on average larger D4000, 
older $\SA$ and higher $\SZ$ than those with smaller $\DSc$ across the probed radial ranges for any given mass and sSFR ranges (Figure \ref{Profile_DS1}), 
which suggests that, similar to T-Types, central stellar mass concentration is a property that is connected with galaxy stellar populations largely in a radius-independent way.

\subsection{Dependence of Stellar Population Profiles on Environments}
\label{sec:env}

\begin{figure*}
   \begin{center}
    \includegraphics[width=1.0\textwidth]{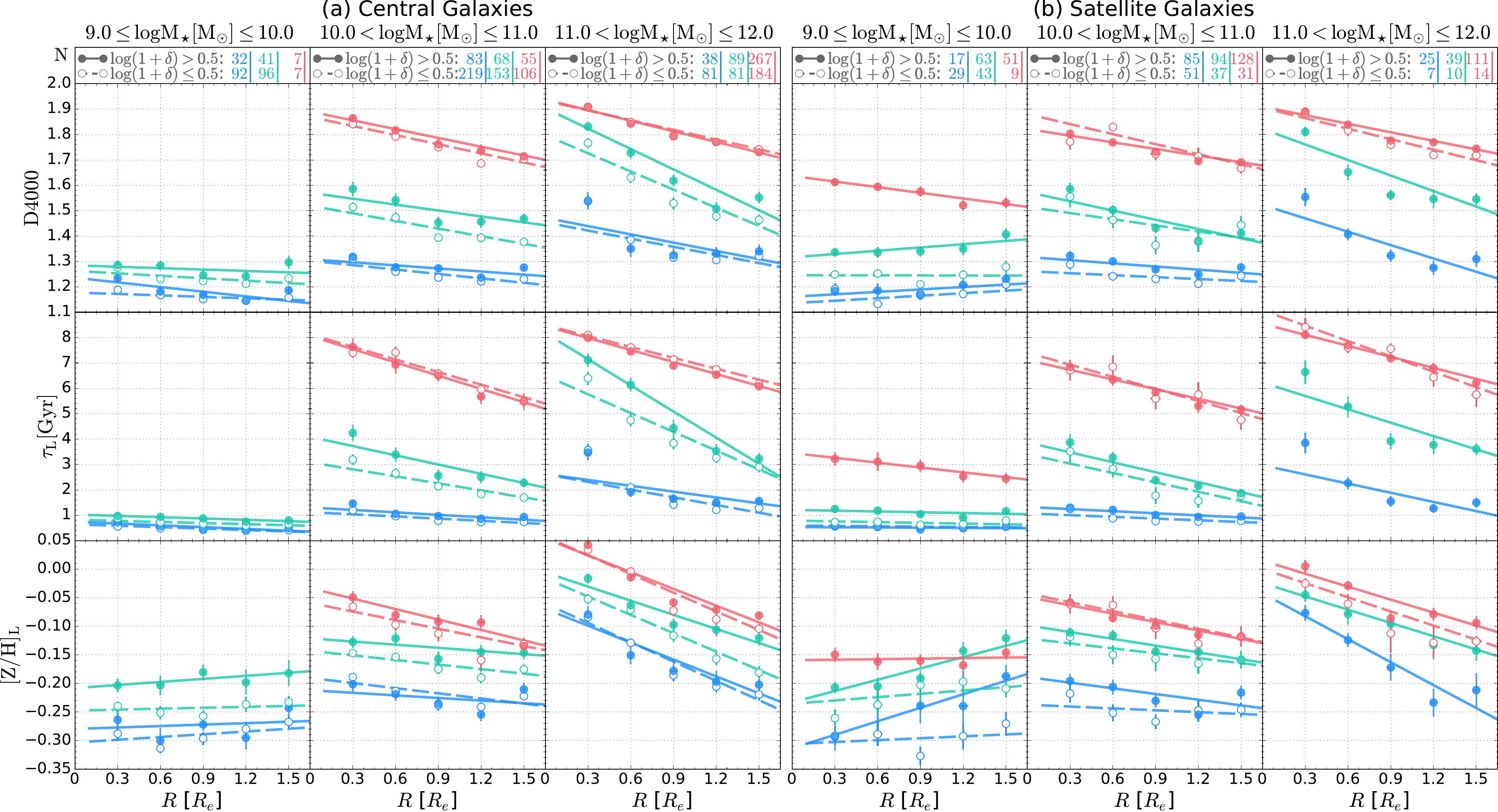}
    \caption{Same as Figure \ref{Profile_TTYPE}, but for the median coadded radial profiles of the D4000, $\SA$, and $\SZ$ of the (a) central galaxies and (b) satellite galaxies located in environments with different local overdensities.\ 
     }
      \label{Profile_LD}
   \end{center}
\end{figure*}

As in Section \ref{sec:mor}, we also explored the dependence of the median $\GD$, $\GA$ and $\GZ$ on local galaxy overdensities $\log (1 + \delta)$ for both central and satellite subsamples, but did not find significant correlations for all but the low-mass star-forming satellite galaxies which exhibit positive correlations between $\GD$ and local overdensities ($\rho \gsim 0.2$) and may indicate a suppression of star formation at larger radii by tidal disturbances of nearby galaxies.\
The corresponding plots are not shown here for clarity.\ The median coadded radial profiles for central and satellite galaxies with $\log (1 + \delta)$ $>$ 0.5 and $\leq 0.5$ are shown in Figure \ref{Profile_LD}.\ For given stellar masses and sSFR, star-forming galaxies located in a higher $\log (1 + \delta)$ environment have slightly (albeit systemically) larger D4000, older $\SA$ and higher $\SZ$ than those located in lower $\log (1 + \delta)$ environment across the probed radial ranges.\ Therefore, the local environment has influence on galaxy stellar populations in a global sense for all star-forming galaxies, and also in a local sense for low-mass star-forming galaxies.

\section{Summary and discussion}
\label{sec:discuss}

\subsection{Summary of major results from this work}

Based on IFU spectroscopic data of a large sample of 3654 nearby galaxies with $10^{9}~\Msun \leq \Ms \leq 10^{12}~\Msun$ from the MaNGA survey available in the SDSS DR15, we have explored the dependence of radial profiles of stellar populations, as quantified by the nearly extinction-free observable D4000, and the 
model-dependent luminosity-weighted stellar ages $\SA$, and luminosity-weighted stellar metallicities $\SZ$, on various galaxy properties.\ We find that $\Ms$ is the most predictive 
global physical property for the D4000 radial gradients $\GD$, and the next most predictive properties for $\GD$, in order of decreasing importance, are sSFR, 
T-Type, and $\DSc$.\ The strongest predictive power of $\Ms$ on $\GD$ reflects the strongest predictive power of $\Ms$ on the metallicity gradient $\GZ$.\ 
The first and second most predictive properties for the age gradient $\GA$ are sSFR and $\Ms$, respectively.\ The local environment quantified by the 
galaxy overdensities $\log (1 + \delta)$ have overall very weak predictive power for radial gradients of the whole sample, but star-forming satellite galaxies 
with $\Ms$ $\lesssim$ $10^{10}~\Msun$ exhibit a positive correlation between $\GD$ and $\log (1 + \delta)$, which may imply a significant suppression of 
star formation activities of relatively low-mass galaxies at large radii by tidal disturbances of nearby galaxies.\

Regarding local stellar population values, the sSFR and, to a much lesser degree, the central velocity dispersion are the most predictive global properties 
for D4000 and $\SA$ values at both the galaxy center and effective radius for our sample.\ For local $\SZ$, $\Ms$ is the most predictive property at the galaxy center, while sSFR is the single most predictive property at one effective radius.\ The environmental properties, as quantified by $\log (1 + \delta)$ and the central-satellite division, have virtually no correlation with galaxy stellar populations for the whole sample, but star-forming galaxies located in a higher $\log (1 + \delta)$ 
environment have on average larger D4000, older $\SA$, and higher $\SZ$ across the probed radial ranges, without regard for being central or satellite.\


The correlation of stellar population gradients with $\Ms$ and sSFR is in the sense that galaxies with higher $\Ms$ or lower sSFR tend to have steeper negative 
gradients.\ Moreover, the negative correlation of median stellar population gradients with $\Ms$ at a given sSFR is best described by segmented relationships, whereby 
galaxies with $\log\Ms \lesssim 10-10.5$ 
(the transition mass being larger for galaxies with larger sSFR) generally have much shallower slopes than those with higher 
$\Ms$.\ Galaxies with a higher global sSFR or younger age generally also have higher sSFR or younger ages locally at given galactocentric distances.\ The morphological 
properties (e.g.,\ T-Type, $\Sc$) are connected with galaxy stellar populations largely in a radius-independent way.\

\subsection{Comparison to relevant results in the literature}

Our finding that the sSFR and, to a much lesser degree, the central velocity dispersion are the most predictive properties for local stellar ages within galaxies is in line with 
a recent study by \citet{Bluck2019}.\ In particular, \citet{Bluck2019} found that the offset from the galaxy SFMS and, to a lesser degree, the central velocity dispersion of host 
galaxies are the two most predictive parameters for whether or not a spatially resolved region is quenched.\ Here we further show that sSFR is also the single most 
predictive global property for $\SZ$ at one effective radius, but at the galaxy center, stellar mass is the most important property for predicting $\SZ$.

The positive correlation between stellar age gradients and the sSFR of star-forming galaxies suggests that (1) quenching of star formation proceeds spatially 
from inside out on average, and this inside-out trend is more significant for more massive galaxies, and (2) the star formation boost with respect to the SFMS primarily 
happens in the inner parts of galaxies, resulting in flatter or even positive stellar population gradients at higher sSFR.\ To the best of our knowledge, previous spectroscopic 
studies of the radial gradients of nearby galaxies mostly focus on the dependence on morphological types and mass (e.g.,\ \citealt{Sanchez-Blazquez2014, GonzalezDelgado2015, 
Zheng2017}), but rarely explore the dependence on the global star formation status of galaxies, which turns out to be the most predictive property for stellar age gradients.\ 
\cite{GonzalezDelgado2015} claimed that it is the Hubble type, rather than stellar mass, that has the closest connection to stellar age gradients, based on 300 
CALIFA galaxies.\ They also found that age/metallicity gradient slopes reach a minimum (i.e.,\ the most negative value) in Sb-Sbc galaxies and increase toward 
either earlier or later Hubble types.\ With an order of magnitude larger sample, we have shown that the sSFR and, to a lesser degree, stellar masses, rather than 
Hubble types (here quantified by T-Types) are the most predictive properties for stellar age gradients.\ At a given sSFR and stellar masses, the radial gradient slopes 
appear to monotonically decrease with T-Types, without an upturn at any intermediate T-Type values.\ Moreover, even at the same Hubble types, stellar age gradients 
generally exhibit a positive dependence on sSFR for given stellar masses (Figure \ref{Grad_TTYPE}).\

The close connection between stellar age gradients and sSFR reported here is also in line with \cite{Ellison2018}, who studied the radial star formation  
profiles of galaxies available in the MaNGA DR13 and found that the relative excess or deficit of star formation as a function of radius is 
correlated with the vertical offset of galaxies with respect to the global SFMS.\ Our analysis suggests that this close connection still exists when dividing 
galaxies into different stellar masses.\ In addition, \cite{Pan2016} studied the radial variations of the broadband (NUV$-r$) color of $\sim$ 6000 local star-forming 
galaxies and found that galaxies with $\log\Ms < 10.2$ have smaller radial color variations and weaker mass dependence than galaxies with higher stellar 
masses.\ Our analysis based on IFU spectroscopy suggests that the stellar mass dependence of radial (NUV$-r$) variations should be primarily attributed to 
a metallicity effect, rather than an age effect.\
\cite{Wang2018a} used the mass-weighted fraction of quenched spaxels with D$_{n}(4000)$ $>$ 1.6 and EW(H$\alpha$) $<$ 2.0 to classify the MaNGA DR14 
galaxies into star-forming, partially quenched (PQ) and totally quenched (TQ) subsamples, and they found that star-forming and PQ galaxies have on average 
similarly negative radial gradients of D$_{n}(4000)$, EW(H$\delta_{A}$), and EW(H$\alpha$), whereas the TQ galaxies have much weaker average gradients 
than star-forming and PQ galaxies over the whole stellar mass ranges.\ This is inconsistent with our finding that stellar population gradients generally become 
more negative as sSFR decreases, except at $\log\Ms \gtrsim 10.5$, where QGs have radial gradients of D4000 and $\SA$ that are shallower than those of 
MS$-$ galaxies but steeper than those of MS$+$ galaxies.\ The difference may be primarily attributed to a difference in the star formation status classification schemes 
adopted in our work and \cite{Wang2018a}.\ The classification of the global star formation status of a galaxy should be based on the globally integrated sSFR, 
regardless of the relative spatial distributions of young and old stellar populations.\ D$_{n}(4000)$ and EW(H$\alpha$) maps reflect the proportional, 
instead of absolute, contribution of younger or poorer-metal stellar populations in any spaxels, which means that spaxels with the same amount of young stellar 
populations may be classified by \cite{Wang2018a} as being star-forming if located in the outer low stellar density regions (or galaxies with lower stellar masses and thus 
generally lower surface mass densities) or quenched if located in the inner high stellar density regions (or galaxies with higher stellar masses and thus higher 
surface mass densities).\ It is thus conceivable that there is some mass-dependent overlap in global sSFRs between the star-forming and PQ subsamples as well as 
between the PQ and TQ subsamples in \cite{Wang2018a}, which tends to smear out the sSFR or mass dependence of stellar population radial gradients. 

By using local stellar mass densities, galactocentric distances (in kiloparsecs) and host galaxy stellar masses as control parameters, \cite{Woo2019} obtained average radial profiles 
of relative enhancement (or offset) of sSFR (or ages) for the relatively isolated galaxies with either compact ($\DSc$ $>$ 0) or diffuse cores ($\DSc$ $<$ 0), and they found that, 
toward smaller galactocentric distances, star-forming galaxies with compact cores are characterized by substantially enhanced median sSFR and depressed chemical abundances,
while those with diffuse cores by substantially suppressed median sSFR and enhanced chemical abundances.\ This led \cite{Woo2019} to conclude that galaxies with $\DSc$ $>$ 0 
follow a compaction-like core-building growth mode whereas galaxies with $\DSc$ $<$ 0 follow a secular inside-out growth mode.\ However, by making a straightforward $R_{e}$-normalized (rather than using physical units of kiloparsecs) radial profile stacking separately for galaxies with different stellar masses and sSFR, we do not observe substantially larger 
$\DSc$-dependent differences between the median D4000/ages at smaller galactocentric distances than at larger galactocentric distances, regardless of the current sSFRs 
(Figure \ref{Profile_DS1}).\ Moreover, star-forming galaxies with compact cores tend to have similar or higher central stellar metallicities than those with relatively diffuse 
cores.\

Our finding of a lack of significant correlation between stellar population gradients and local galaxy overdensities is largely in line with previous studies based on MaNGA 
data \citep{Goddard2017a,Zheng2017,Spindler2018}.\ Nevertheless, we additionally show that this lack of correlation applies to both star-forming and quiescent galaxies, 
which implies that local overdensities do not significantly affect the stellar population distribution in galaxies over either short or long timescales.\

\subsection{Implications for a Negative Dependence of Stellar Population Gradients on Stellar Mass}

The classical monolithic dissipative collapse models of galaxy formation (e.g., \citealt{Larson1974, Carlberg1984, Pipino2010}) are qualitatively consistent 
with our finding of a monolithic negative dependence of stellar metallicity gradients on stellar mass, but are inconsistent with the overall negative stellar 
age gradients.\ In the $\Lambda$CDM hierarchical structure formation models, galaxy mergers are expected to be common especially in the early 
universe \citep[e.g.,][]{Duncan2019}.\ The early gas-dominated merging events (along with accretion-driven violent disk instabilities; e.g., \citealt{Dekel_Burkert_2014})
might have formed the central parts of the present-day galaxies first, and later, the fallback of relatively high angular momentum cold gas ejected during the merging 
process might have rebuilt extended disks from inside out (e.g., \citealt{Barnes2002}), giving rise to negative radial gradients of stellar ages and metallicities.\ 
Some numerical simulations \citep[e.g.,][]{Bekki1999,Hopkins2009} suggest that gas-rich major mergers involving more massive progenitors give rise to steeper 
metallicity gradients due to a stronger central starburst sustained by tidally induced gas inflow.\

Nevertheless, more recent simulations by \cite{Taylor2017} suggest that gas-rich major mergers do not significantly steepen metallicity gradients 
when active galactic nucleus (AGN) feedback is invoked to suppress the central SF activities.\ Instead, the redistribution of high-metallicity stars from the center to outskirts caused by
merging tends to flatten preexisting metallicity gradients, resulting in a reversal of the mass dependence of metallicity gradients from being negative at 
$\log\Ms \lesssim 10.5$ to positive or flat at higher stellar masses.\ A similar reversal of metallicity gradients was also found by \cite{Tissera2016} in their 
simulation of disk galaxies, even though AGN feedback was not included.\ Moreover, \cite{Tissera2016} did not find a clear correlation between age gradients 
and stellar mass.\ These recent simulation results are in conflict with our observational finding of a monolithic mass dependence of the average age  
and particularly metallicity gradients.\ Nevertheless, we emphasize that the analysis presented in this work is limited to the central $\sim$ 1.5$R_{e}$ 
of nearby galaxies.\ Studies extending to larger galactocentric distances indeed found that age and metallicity gradients of nearby galaxies may flatten beyond 
$2$$R_{e}$ \citep[e.g.,][]{GonzalezDelgado2015}, indicating that either satellite accretion or radial stellar migration might play an important role in the outer disks.





\acknowledgments

This work is supported by the National Key R\&D Program
of China (2017YFA0402600, 2017YFA0402702), the B-type
Strategic Priority Program of the Chinese Academy of Sciences
(XDB41000000), and NSFC grant Nos.\ 11973038, 11973039,
and 11421303.\ H.X.Z. is also grateful for support from the
CAS Pioneer Hundred Talents Program.


This work makes use of the MaNGA-Pipe3D dataproducts. We thank the IA-UNAM MaNGA team for creating this catalogue, and the Conacyt Project CB-285080 
for supporting them.\ Funding for the Sloan Digital Sky Survey IV has been provided by the Alfred P. Sloan Foundation, the U.S. Department of Energy Office of 
Science, and the Participating Institutions. SDSS-IV acknowledges support and resources from the Center for High-Performance Computing at the University of 
Utah. The SDSS web site is www.sdss.org.

\bibliographystyle{aasjournal}

\bibliography{grad}

\restartappendixnumbering

\appendix

\section{Results for mass-weighted stellar ages and metallicities}
\begin{figure*}[ht!]
   \begin{center}
   \includegraphics[width=1.0\textwidth]{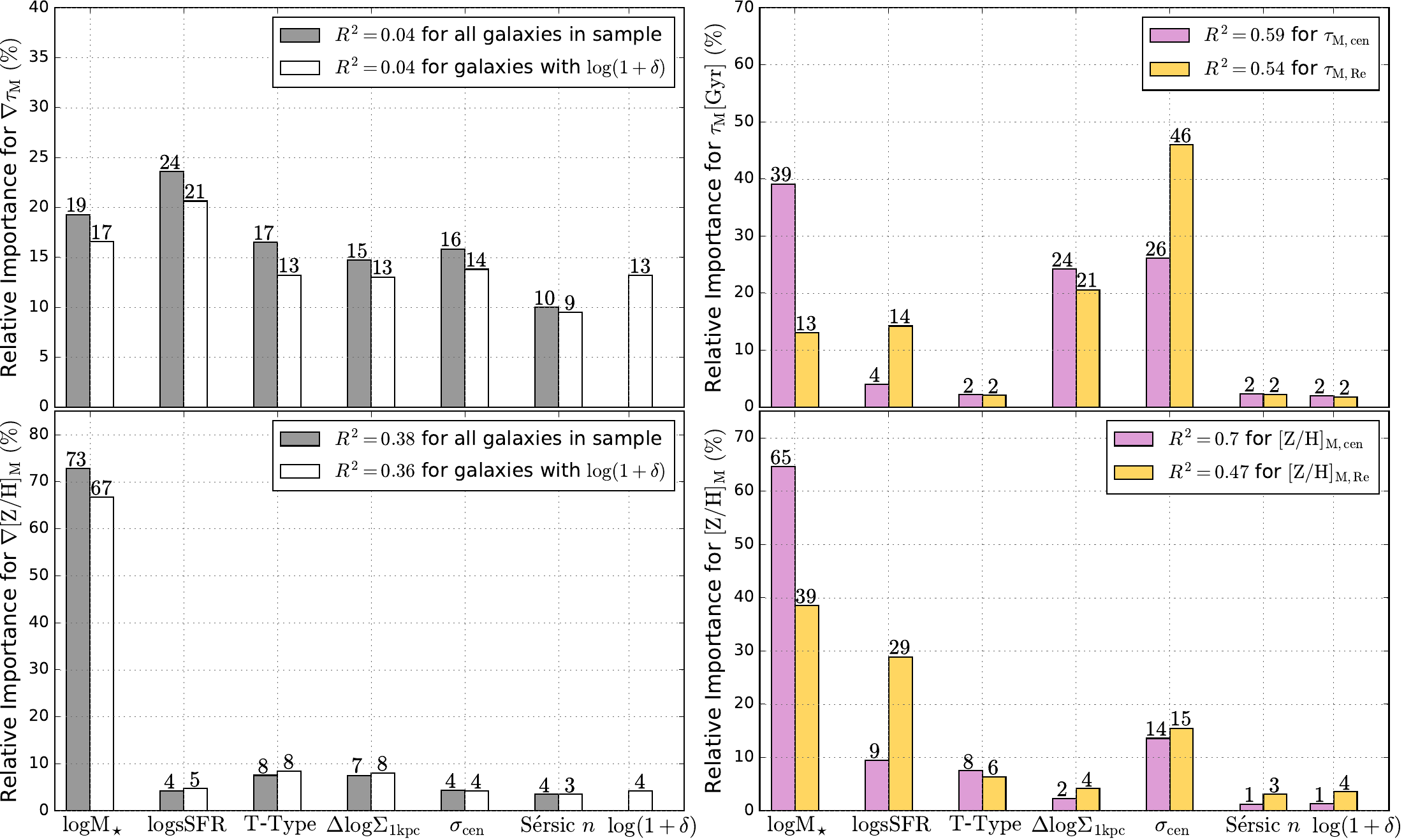}
      \caption{
         Same as Figure \ref{RFimportance}, but for the evaluation of the relative importance of different galaxy properties in predicting the radial gradients of mass-weighted ages and metallicities ($\GAM$ and $\GZM$), and mass-weighted ages $\SAM$ and metallicities $\SZM$ at galaxy centers (purple bars) and $\Reff$ (orange bars).
      } 
      \label{RFimportance_MW}
   \end{center}
\end{figure*}

\begin{figure*}[ht!]
   \begin{center}
   \includegraphics[width=1.0\textwidth]{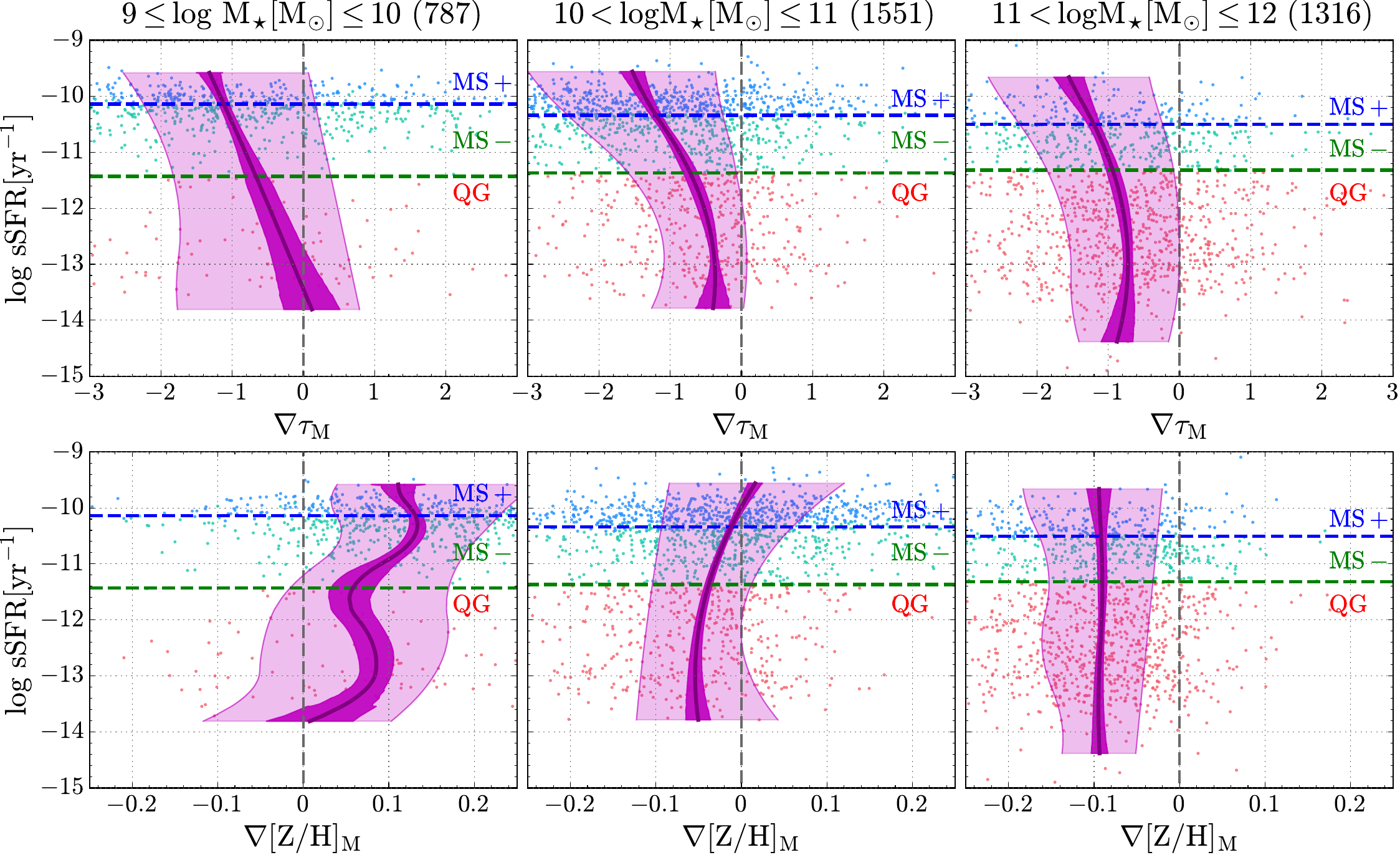}
      \caption{
         Same as Figure \ref{Grad}, but for global sSFR plotted against mass-weighted stellar population gradients ($\GAM$ and $\GZM$) for galaxies in different $\log\rm{M_\star}$ bins.
      }
      \label{Grad_MW}
   \end{center}
\end{figure*}

\begin{figure*}[ht!]
   \begin{center}
   \includegraphics[width=0.75\textwidth]{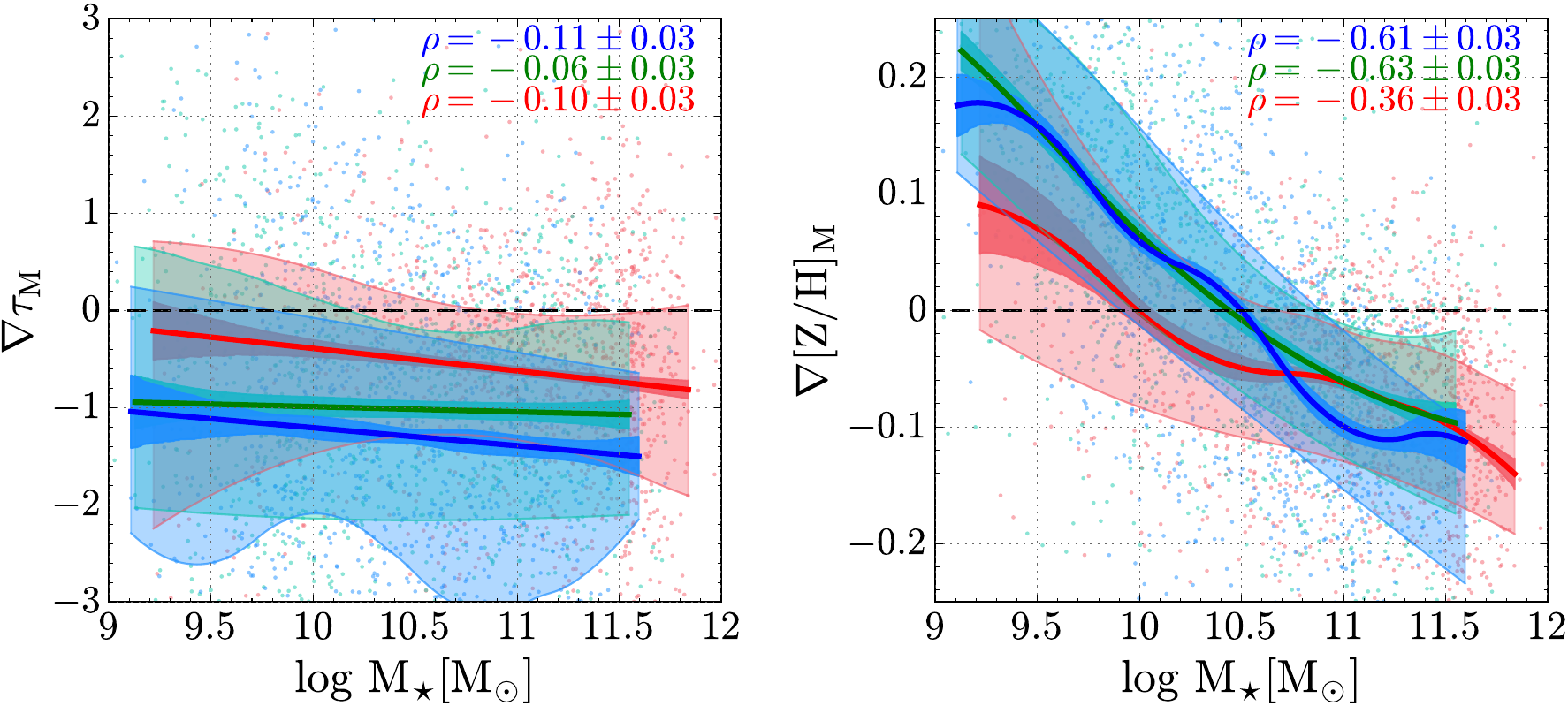}
      \caption{
      Same as Figure \ref{Grad_Mass}, but for 
      mass-weighted stellar age and metallicity gradients (from left to right: $\GAM$ and $\GZM$) as a function of stellar masses for MS$+$ (blue), MS$-$ (green), and QG (red).\ 
      }
      \label{Grad_Mass_MW}
   \end{center}
\end{figure*}

While we have restricted our discussion of model-dependent stellar parameters to the luminosity-weighted ages and metallicities in the main text of this paper, here we show the relevant results for mass-weighted ages and metallicities for the sake of completeness (Figures \ref{RFimportance_MW}, \ref{Grad_MW}, and \ref{Grad_Mass_MW}).\ A brief discussion about the problem with mass-weighted stellar parameters derived based on integrated spectra is given in Section \ref{sec:meth}.




\end{document}